\documentclass[aps,pre,superscriptaddress, preprint]{revtex4-1}
\usepackage[utf8]{inputenc}
\usepackage{amsmath, amssymb,stmaryrd}
\usepackage{graphicx}
\usepackage[braket, qm]{qcircuit}
\usepackage{slashed}
\begin{document}

\title{Scalable Qubit Representations of Neutrino Mixing Matrices}
\author{M. J. Molewski}
\thanks{\url{matthew.molewski@mavs.uta.edu}}
\affiliation{Department of Physics, University of Texas at Arlington, Arlington, TX 76019, USA}
\author{B. J. P. Jones}
\thanks{\url{ben.jones@uta.edu}}
\affiliation{Department of Physics, University of Texas at Arlington, Arlington, TX 76019, USA}

\begin{abstract}
    Oscillating neutrino beams exhibit quantum coherence over distances of thousands of kilometers.  Their unambiguously quantum nature suggests an appealing test system for direct quantum simulation. Such techniques may enable presently analytically intractable calculations involving multi-neutrino entanglements, such as collective neutrino oscillations in supernovae, but only once oscillation phenomenology is properly re-expressed in the language of quantum circuits.  Here we resolve outstanding conceptual issues regarding encoding of arbitrarily mixed neutrino flavor states in the Hilbert space of an n-qubit quantum computer.  We introduce algorithms to encode mixing and oscillation of any number of flavor-mixed neutrinos, both with and without CP-violation, with an efficient number of prescriptive input parameters in terms of sub-rotations of the PMNS matrix in standard form.  Examples encoded for an IBM-Q quantum computer are shown to converge to analytic predictions both with and without CP-violation.

\end{abstract}
\maketitle

\section{Introduction}

Since first theorized in 1930 by Pauli~\cite{brown1978idea}, neutrinos have played a truly fundamental role in understanding the structure of the Standard Model of particle physics~\cite{reines1956neutrino}.  Initially introduced to solve the problem of energy non-conservation in beta decay, the neutrino was initially assumed to be massless, a postulate that remains consistent with kinematic measurements even to this day for the lightest neutrino mass state~\cite{aker2019improved}.  The discovery of neutrino oscillations~\cite{giunti2007fundamentals}, however, demonstrated both a nonzero mass for at least two of the three neutrino states, and also that neutrino production via weak interactions produces quantum superpositions of these mass basis states. Taking inspiration from the Cabibbo–Kobayashi–Maskawa (CKM) matrix used to describe quark mixing, Pontecorvo, Maki, Nakagawa and Sakata proposed that a basis misalignment between a neutrino's flavor and mass eigenstates and subsequent evolution in time could be described using what would become known as the Pontecorvo-Maki-Nakagawa-Sakata matrix ($U_{PMNS}$) \cite{maki1962remarks},
\begin{equation}
    |\nu_{\alpha}\rangle=U_{PMNS}|m_i\rangle ,
\end{equation}
with $\alpha=e,\mu,\tau$ labeling neutrino flavor and $i=0,1,2$ enumerating the neutrino mass states.  In models with a fourth, sterile neutrino~\cite{Kopp:2013vaa}, an additional mass state $i=3$ is added, along with a corresponding sterile flavor state $\alpha=s$.  Given a PMNS matrix, the probability for neutrino oscillations from flavor $\alpha$ to $\beta$ is described (in vacuum) by,
\begin{equation}
    P_{\alpha\rightarrow\beta}=\left|\sum_i U_{\alpha i}^*U_{\beta i}e^{-i\frac{m_i^2 L}{2E}} \right|^2.
\end{equation}
With $L$ the baseline of neutrino propagation, $E$ the neutrino energy, and $m_i$ the mass of the $i$th neutrino mass state. Since $U_{PMNS}$ is a three dimensional unitary matrix, following arbitrary phase choices of the neutrino flavor and mass fields, it can be parametrized in the following conventional way in terms of three ``mixing angles'' $\theta_{13}$, $\theta_{23}$ and $\theta_{12}$, and one CP-violating phase $\delta_{CP}$ for non-Majorana neutrinos, as
\begin{equation}    
 U_{PMNS}= \begin{bmatrix} 1 & 0 & 0\\ 0 & c_{23} & s_{23}\\ 0& -s_{23} & c_{23}\end{bmatrix} \begin{bmatrix} c_{13} & 0 & s_{13}e^{-i\delta_{CP}}\\ 0 & 1 & 0\\ -s_{13}e^{i\delta_{CP}}& 0 & c_{13}\end{bmatrix} \begin{bmatrix} c_{12} & s_{12} & 0\\ -s_{12} & c_{12} & 0\\ 0& 0 & 1\end{bmatrix}\label{eq:PMNSStandsard}, 
\end{equation}
where here $c_{ij}$ and $s_{ij}$ refer to the cosine and sine, respectively, of $\theta_{ij}$. This can be conveniently written as,
\begin{equation}
    U_{PMNS}=R^{23}(\theta_{23})R^{13}(\theta_{13},\delta_{CP})R^{12}(\theta_{12})\label{eq:PMNS},
\end{equation}
where $R^{ij}(\theta,\delta)$ represents a rotation in the $ij$ plane with angle $\theta$ and a phase rotation $\delta$. We have used notation such that $R^{ij}(\theta)\equiv R^{ij}(\theta,0)$.  

Direct quantum simulation of neutrino oscillations was first  introduced in Ref.~\cite{arguelles2019neutrino} as an example of a simple and illustrative problem amenable to application on few-qubit publicly available quantum computers. Quantum simulation, as popularized by Feynman~\cite{feynman2018simulating} and others~\cite{manin1980computable,benioff1980computer}, involves encoding the dynamics of a system under study in the Hilbert space of a quantum processor in order to solve for its time evolution with application of an appropriately encoded time evolution operator.  In future highly entangled quantum computers, this may offer opportunities for efficient evaluation of calculations that are classically intractable. Circuits and algorithms have been proposed or implemented to model processes in chemistry~\cite{kassal2011simulating}, biology~\cite{warshel2001nature}, neural networks~\cite{schuld2015simulating}, pharmacology~\cite{mulligan2020designing}, quantum gravity~\cite{mielczarek2019spin}, quantum chaos~\cite{schack1998using} and quantum chromodynamics~\cite{atas20212}.  

Neutrino oscillations in vacuum are a fairly simple quantum system and simple closed-form analytic solutions to their time evolution are easy to obtain.  In other conditions neutrino oscillations become much more difficult to model. Two examples where the calculation of neutrino time evolution becomes far more complex include the propagation of high energy neutrinos in matter~\cite{smirnov2005msw,delgado2015simple} and the evolution of high neutrino density ensembles in supernova explosion exhibiting collective oscillation effects~\cite{duan2010collective,duan2006collective}.  The latter has very recently been the subject of attempts at quantum simulation~\cite{yeter2021collective,hall2021simulation}, though only in a simplified  two-flavor basis. In such a basis, one qubit maps directly onto one neutrino via the association $|\nu_e\rangle\rightarrow|0\rangle$ and $|\nu_X\rangle\rightarrow|1\rangle$, where X is a composite flavor representing a generic non-electron neutrino.

An important technical contribution of Ref.~\cite{arguelles2019neutrino} was the demonstration of an encoding of the three-neutrino Hilbert space onto two qubits, which required a non-trivial mapping of the PMNS mixing matrix, in addition to the much simpler mapping of the mass-basis Hamiltonian, onto this space via the available quantum gates of the IBM-Q system. IBM-Q is a publicly available universal gate quantum computer~\cite{santos2016ibm} to which quantum circuit calculations can be submitted via web interface~\cite{IBMQ}.  

Since it represents a simple rotation in two dimensions, the PMNS matrix for a two-flavor neutrino system has a nearly trivial form,
$$\scalebox{1.0}{\Qcircuit @C=1.0em @R=0.2em @!R {
	 	\nghost{ {q}_{0} : } & \lstick{ {q}_{0} :  } & \gate{\mathrm{U_3}\,(\mathrm{2\theta_{12},0,0})} & \qw & \qw  }} \ ,$$
where $\theta_{12}$ represents the mixing angle between the electron and muon neutrino flavors. The U3 gate is defined as,
$$U3(\theta,\phi,\lambda)=\begin{bmatrix}
\cos(\theta/2) & -\sin(\theta/2)e^{i\lambda}\\ \sin(\theta/2)e^{i\phi} & \cos(\theta/2)e^{i(\phi+\lambda)}
\end{bmatrix} . $$
For the 3-flavor system, construction of the PMNS matrix is more involved. In Ref.~\cite{arguelles2019neutrino}  this was accomplished for the special case of $\delta_{CP}=0$ by first constructing a system of six U3 gates and 2 CNOT operators, as,
$$\scalebox{1.0}{
\Qcircuit @C=1.0em @R=0.2em @!R { \\
	 	\nghost{ {q}_{0} :  } & \lstick{ {q}_{0} :  } & \gate{\mathrm{U_3}\,(\mathrm{\alpha,0,0})} & \targ & \gate{\mathrm{U_3}\,(\mathrm{\gamma,0,0})} & \targ & \gate{\mathrm{U_3}\,(\mathrm{\epsilon,0,0})} & \qw & \qw\\ 
	 	\nghost{ {q}_{1} :  } & \lstick{ {q}_{1} :  } & \gate{\mathrm{U_3}\,(\mathrm{\beta,0,0})} & \ctrl{-1} & \gate{\mathrm{U_3}\,(\mathrm{\delta,0,0})} & \ctrl{-1} & \gate{\mathrm{U_3}\,(\mathrm{\zeta,0,0})} & \qw & \qw\\ 
\\ }} \ , $$
where the basis is defined as $|00\rangle$ the electron neutrino, $|01\rangle$ the muon neutrino, $|10\rangle$ the tau neutrino and $|11\rangle$ an unphysical sterile neutrino.  The free parameters were then obtained by fitting to mixing angles obtained from world neutrino oscillation data~\cite{particle2020review} using gradient descent minimization.  While analytic connection between the gate parameters and conventional neutrino mixing angles was not made in that work, direct substitution can be used to obtain relations between the circuit parameters and the entries of the PMNS matrix, yielding, 
\begin{eqnarray}
\frac{s_{12}c_{13}}{2\cos(\delta/2)}= \\ 
-\sin(\beta/2)\cos(\zeta/2)\cos[(\alpha+\gamma+\epsilon)/2]-\cos(\beta/2)\sin(\zeta/2)\cos[(\epsilon+\alpha-\gamma)]; \nonumber \\ 
\frac{-s_{12}c_{23}-c_{12}s_{23}s_{13}}{2\sin(\delta/2)}= \\ 
\sin(\beta/2)\sin(\zeta/2)\sin[(\epsilon+\gamma-\alpha)/2]-\cos(\beta/2)\cos(\zeta/2)\sin[(\epsilon-\alpha-\gamma)]; \nonumber \\
\frac{c_{12}c_{23}-s_{12}s_{23}s_{13}+1}{2\cos(\delta/2)}= \\ 
-\sin(\beta/2)\sin(\zeta/2)\cos[(\alpha+\gamma+\epsilon)/2]+\cos(\beta/2)\cos(\zeta/2)\cos[(\epsilon+\alpha-\gamma)], \nonumber 
\end{eqnarray}
from which appropriate values for the circuit parameters can be established, given known values of the PMNS matrix elements.  However, the approach of Ref.~\cite{arguelles2019neutrino}, which was co-authored by one of us, left important aspects of this problem incompletely solved.  In particular: 1) The form of the circuit was obtained by trial and error, following attempts to map to the PMNS operator after parameter fitting. Any physical insight about its form in terms of neutrino mixing angles has therefore been lost. 2) The parametrization is clearly not minimal since six free parameters are used to generate a real operator which is known to be expressible in terms of three parameters; 3) The circuit cannot accommodate a non-zero value for the CP-violating phase; 4) There is not an obvious way to extend this approach to higher dimensionality or to efficient encoding of multi-neutrino states.  

In this paper we resolve all of these issues.  We derive a new circuit to encode the three dimensional PMNS matrix that uses a minimal set of free parameters and does not require any numerical fitting to fix their values.  Our new approach provides an intuitive mapping between the three-neutrino and two-qubit Hilbert spaces, and also allows for simple incorporation of the CP violating phase. We then extend this method to construction of arbitrarily high dimensional PMNS operators. A four dimensional extension on two qubits is straightforwardly derived. Performance of these circuits is tested on IBM-Q and accuracy comparable with the circuits of Ref~\cite{arguelles2019neutrino} are recovered.

\section{Notation and Standard Gates}


In this work we will make repeated use of general controlled and anti-controlled U3 gates.  Our notation will be such that an operator that is controlled on qubit $i$ and applies operator $U3$ to target qubits $abc$ will be represented as $CU^i_{abc}$, and an anti-controlled operator that is anti-controlled on qubit $i$ and applies operator $U3$ to target qubits $abc$ will be represented as $\slashed CU3^i_{abc}$.  An operator that is controlled on qubit $i$ and acts on all other qubits will be represented $CU^i_{[...]}$.

The Hilbert space will be described either in terms of qubit values running from $|000...\rangle$ to $|111...\rangle$ with the most significant bit on the left, or in terms of basis states numbered from $|1\rangle$ to $|2^N\rangle$. The qubit representation related to the state number representation by subtracting 1 and converting to binary.

As an example, in the two qubit system the basis can be expressed as either $[
|00\rangle,  |01\rangle, |10\rangle, |11\rangle]$ or $[
|1\rangle,  |2\rangle, |3\rangle, |4\rangle]$, and the following are the complete set of controlled and anticontrolled $U3$ gates,
\begin{equation}
    CU3^1_0(\theta, \phi, \lambda)=\begin{bmatrix} 1 &0 & 0 &0\\ 0& \cos(\theta/2) & 0 & -\sin(\theta/2)e^{i\lambda} \\ 0 & 0 & 1 &0\\ 0 & \sin(\theta/2)e^{i\phi} & 0 & \cos(\theta/2)e^{i(\phi+\lambda)}\end{bmatrix};
\end{equation}
\begin{equation}CU3^0_1(\theta, \phi, \lambda)=\begin{bmatrix} 1 &0 & 0 &0\\ 0& 1 & 0 & 0 \\ 0 & 0 & \cos(\theta/2) &-\sin(\theta/2)e^{i\lambda}\\ 0 & 0 & \sin(\theta/2)e^{i\phi} & \cos(\theta/2)e^{i(\phi+\lambda)}. \end{bmatrix};
\end{equation}

\begin{equation}
\slashed{C}U3^1_0(\theta, \phi, \lambda)=\begin{bmatrix} \cos(\theta/2) &0 & -\sin(\theta/2)e^{i\lambda} &0\\ 0& 1 & 0 & 0  \\ \sin(\theta/2)e^{i\phi} & 0 & \cos(\theta/2)e^{i(\phi+\lambda)} &0\\ 0 & 0 & 0 & 1\end{bmatrix};
\end{equation}
\begin{equation}
\slashed{C}U3^0_1(\theta,\phi, \lambda)=\begin{bmatrix} \cos(\theta/2) &-\sin(\theta/2)e^{i\lambda} & 0 &0\\ \sin(\theta/2)e^{i\phi}& \cos(\theta/2)e^{i(\phi+\lambda)} & 0 & 0 \\ 0 & 0 & 1 &0\\ 0 & 0 &  0& 1 \end{bmatrix}. \end{equation}
Note that the controlled and anticontrolled operators are related to one another by the transformation,
\begin{equation}
\scalebox{1.0}{
\Qcircuit @C=1.0em @R=0.2em @!R {
	 	\nghost{ {q}_{0} :  } & \lstick{ {q}_{0} :  } & \ctrlo{1} & \qw & \qw\\ 
	 	\nghost{ {q}_{1} :  } & \lstick{ {q}_{1} :  } & \gate{\mathrm{U_3}\,(\mathrm{-2\theta,\delta,-\delta})} & \qw & \qw\\ 
\\} \quad { \\ =} \Qcircuit @C=1.0em @R=0.2em @!R { 
	 	\nghost{ {q}_{0} :  } & \lstick{ {q}_{0} :  } & \gate{\mathrm{X}} & \ctrl{1} & \gate{\mathrm{X}} & \qw & \qw\\ 
	 	\nghost{ {q}_{1} :  } & \lstick{ {q}_{1} :  } & \qw & \gate{\mathrm{U_3}\,(\mathrm{-2\theta,\delta,-\delta})} & \qw & \qw & \qw\\  }}.
\end{equation}
so that, even though our algorithms make liberal use of anticontrol bits to simplify notation, they can still be straightforwardly implemented on quantum architectures such as IBM-Q which do not feature anticontrolled gates in their default tool set.

\section{Three- and four-dimensional PMNS matrices on a two-qubit Hilbert space~\label{sec:PMNS3D}}

Our approach to intuitive construction of the PMNS operator from physical arguments will be to start with the standard parameterization of Eq.~\ref{eq:PMNSStandsard} and seek expressions for each of the sub-rotations with quantum circuits.  The protocol we develop will then be generalized to systems of arbitrarily high dimensionality.

For each case we will begin with one chosen ``base sub-rotation,'' by convention $R^{12}(\theta_{12})$, though any other choice would work equally well.  By inspection of the two-qubit anti-controlled U3 operator in the neutrino Hilbert space we observe that this operator can be represented as,
\begin{equation}
  R^{12}(\theta,\delta)=\slashed{C}U3^0_1(-2\theta, \delta, -\delta),
\end{equation}
or in circuit notation,
\begin{equation}
\scalebox{1.0}{
\Qcircuit @C=1.0em @R=0.2em @!R {
	 	\nghost{ {q}_{0} :  } & \lstick{ {q}_{0} :  } & \ctrlo{1} & \qw & \qw\\ 
	 	\nghost{ {q}_{1} :  } & \lstick{ {q}_{1} :  } & \gate{\mathrm{U_3}\,(\mathrm{-2\theta,\delta,-\delta})} & \qw & \qw\\ 
	 	}}
\end{equation}
With this base rotation in hand, we note that any other can be found by applying a suitable basis transformation, in particular a matrix permutation operation.  We denote a permutation that swaps the rows ${i,j}$ in a $2^n+1$ dimensional Hilbert space on $n$ qubits as $(ij;\mathbf{n})$. If each $(ij,\mathbf{n})$ can be found in terms of available quantum gates, any sub-rotation can be obtained from the base sub-rotation via,
\begin{eqnarray}
  R^{ij}(\theta,\delta)&=&(i1;\mathrm{n})(j2;\mathrm{n})R^{12}(\theta,\delta)(i1;\mathrm{n})^{-1}(j2;\mathrm{n})^{-1}\nonumber\\
  &=&(i1;\mathrm{n})(j2;\mathrm{n})R^{12}(\theta,\delta)(i1;\mathrm{n})(j2;\mathrm{n}).\label{eq:BaseSubrotPermut}
\end{eqnarray}
Finding the full set of required sub-rotations is then reduced to the problem of finding and applying appropriate permutation matrices to the base sub-rotation. In two qubit systems the required permutations can be obtained by inspection of the matrix form of the operator on the computational Hilbert space. For example, to find the $R^{13}$ matrix we require the $(23;\mathbf{2})$ permutation, which we can identify by inspection as the 2-qubit SWAP operation,
\begin{equation}
(23;\mathbf{2})=\begin{bmatrix} 1 &0 & 0 &0\\ 0& 0 & 1 & 0 \\ 0 & 1 & 0 &0\\ 0 & 0 &0 & 1 \end{bmatrix}=SWAP.
\end{equation}
And so the circuit representation of $R^{13}$ is obtained, as,
\begin{equation}
R^{13}(\theta,\delta)=(SWAP)R^{12}(\theta,\delta)(SWAP).
\end{equation}
or in circuit notation,
\begin{equation}
\scalebox{1.0}{
\Qcircuit @C=1.0em @R=0.2em @!R {
	 	\nghost{ {q}_{0} :  } & \lstick{ {q}_{0} :  } & \qswap &  \ctrlo{1} & \qswap & \qw & \qw\\ 
	 	\nghost{ {q}_{1} :  } & \lstick{ {q}_{1} :  } & \qswap \qwx[-1] &  \gate{\mathrm{U_3}\,(\mathrm{-2\theta,\delta,-\delta})} &  \qswap \qwx[-1] & \qw & \qw\\ 
}}
\end{equation}
\begin{figure}[t]
    \centering
    \includegraphics[width=0.8\textwidth]{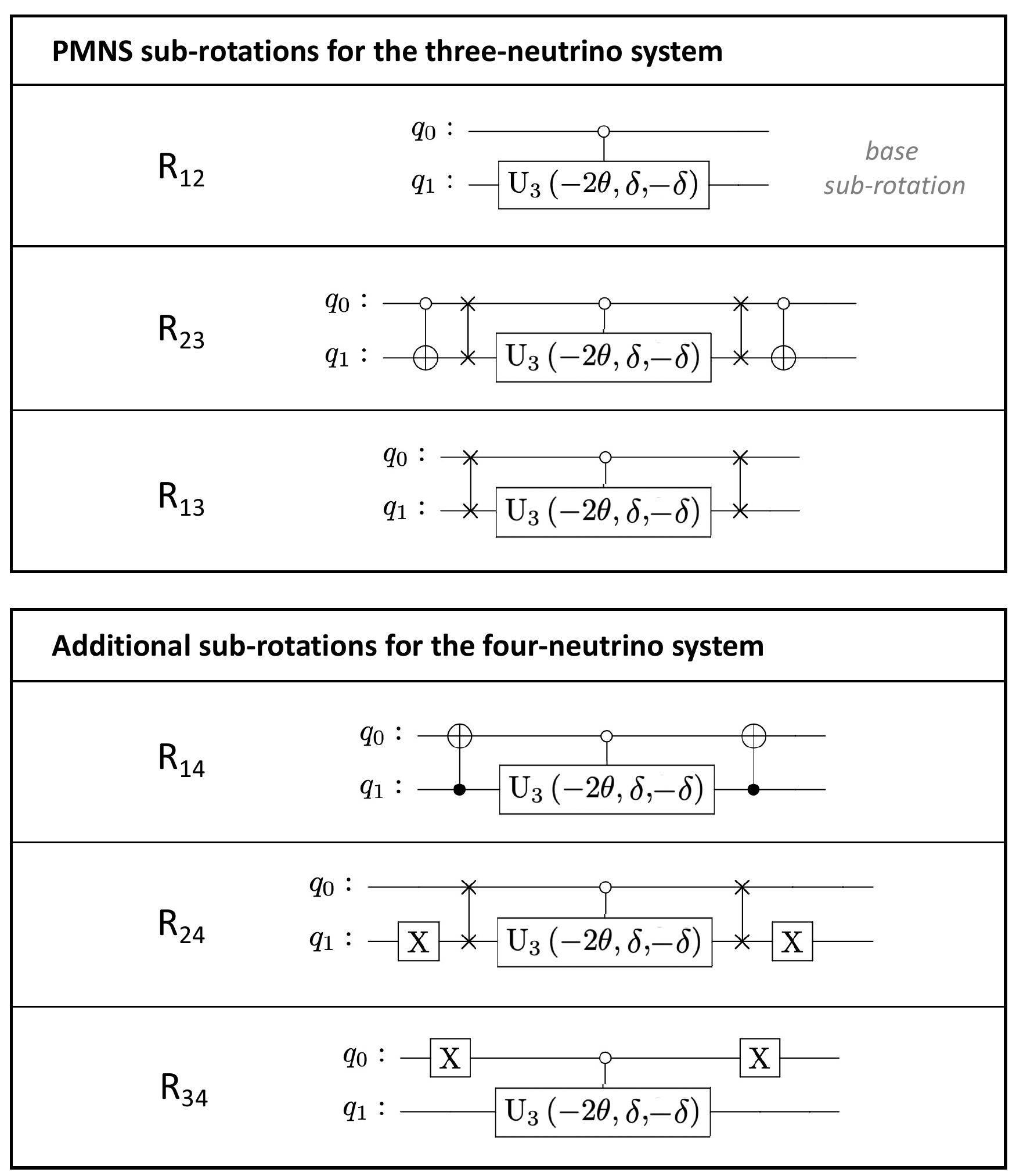} 
    \caption{The full set of sub-rotations needed to realize the PMNS matrix in three- and four-neutrino systems on a two-qubit quantum computer, generated via permutations of the $R_{12}$ base sub-rotation.}
    \label{fig:SubRotFigure}
\end{figure}
In this manner all the sub-rotations required for the two-qubit system can be obtained.  The full set required for simulating three- and four-neutrino systems on a two-qubit quantum computer are given in Fig.~\ref{fig:SubRotFigure}.

For a computation involving only the three known active neutrinos, only $R_{12}(\theta_{12},\delta=0)$, $R_{23}(\theta_{23},\delta=0)$ and $R_{13}(\theta_{13},\delta_{CP})$ are required. Connecting together these circuit elements and performing some simple cancellations and gate manipulations, we arrive at the three dimensional PMNS matrix, using Eq.~\ref{eq:PMNS},
\begin{equation} 
\scalebox{1.0}{
\Qcircuit @C=1.0em @R=0.2em @!R { \\
	 	\nghost{ {q}_{0} :  } & \lstick{ {q}_{0} :  } & \ctrlo{1} & \gate{\mathrm{U_3}\,(\mathrm{-2\theta_{13},\delta_{CP},-\delta_{CP}})} & \targ & \ctrlo{1} & \targ & \qw & \qw\\ 
	 	\nghost{ {q}_{1} :  } & \lstick{ {q}_{1} :  } & \gate{\mathrm{U_3}\,(\mathrm{-2\theta_{12},0,0})} & \ctrlo{-1} & \ctrlo{-1} & \gate{\mathrm{U_3}\,(\mathrm{2\theta_{23},0,0})} & \ctrlo{-1} & \qw & \qw\\ 
\\ }}
\end{equation}
In a system with a fourth, sterile neutrino state participating in oscillations, a similar circuit can be generated via,
\begin{equation}
    \mathrm{PMNS}_{4\times4}=R^{34}(\theta_{34})R^{24}(\theta_{24},\delta_{24})R^{14}(\theta_{14},\delta_{14})R^{23}(\theta_{23})R^{13}(\theta_{13},\delta_{CP})R^{12}(\theta_{12}).
\end{equation}
We test predictions of the oscillation circuit formed from such two-qubit PMNS constructions on both a quantum simulator and a quantum computer in the subsequent section.

\section{Evaluation of three-neutrino oscillations with and without CP violation on the IBM-Q system}

\begin{figure}[t]
    \centering
    \includegraphics[width=0.49\textwidth]{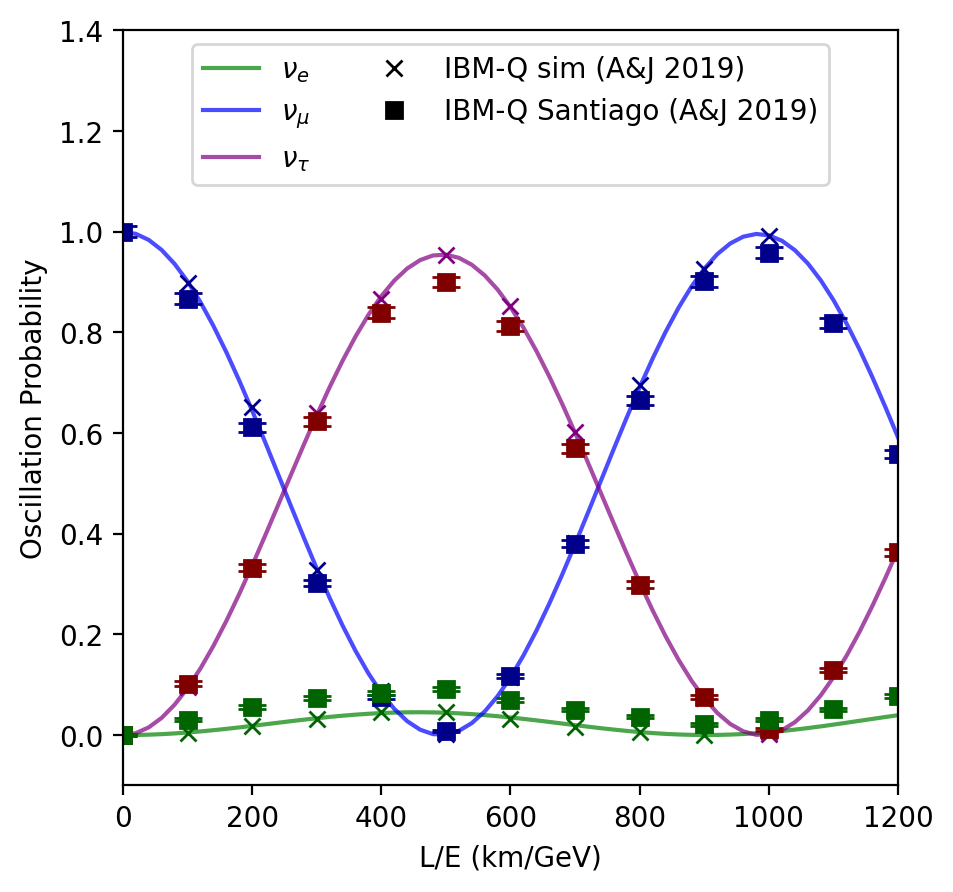} 
    \includegraphics[width=0.49\textwidth]{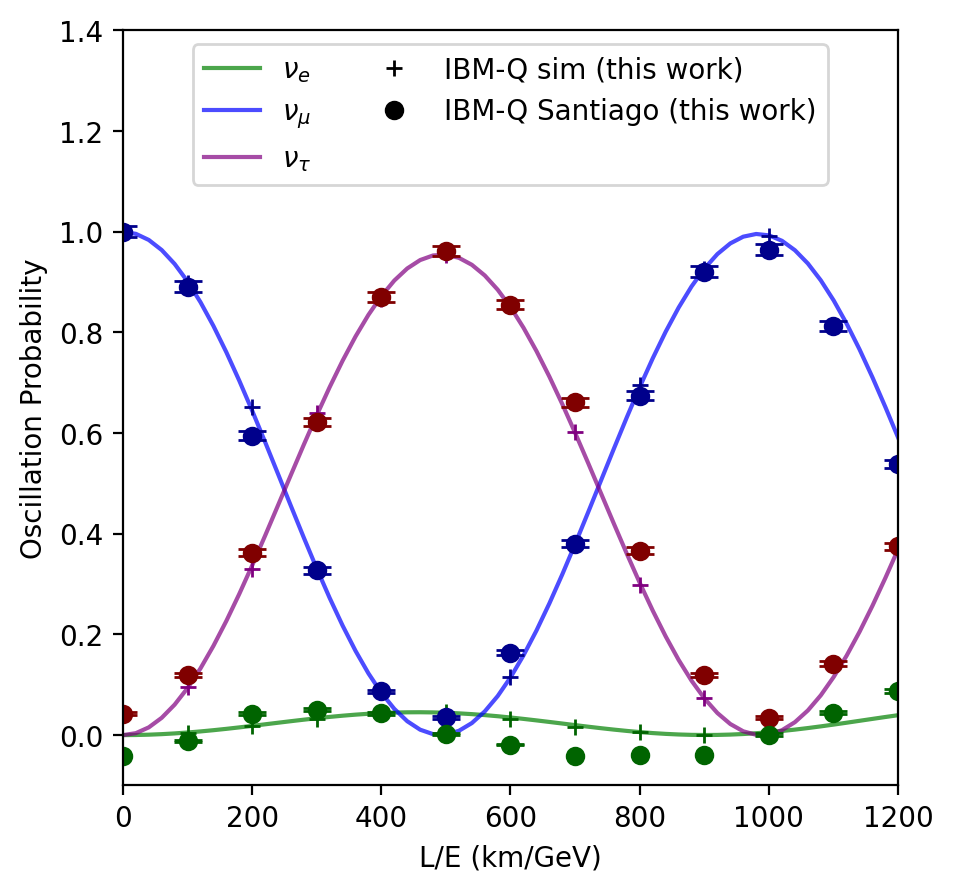} 
    \caption{Calculations on the IBM-Q QASM simulator and {\tt IBM-Q Santiago} quantum computer using the 3-flavor PMNS circuits discussed in this work, without CP violation.  Left shows results of the original empirical circuit of Ref.~\cite{arguelles2019neutrino} reproduced on the new processor; Right shows results of the new physically motivated circuit developed in this work.}
    \label{fig:ThreeFlavor}
\end{figure}

The PMNS circuit developed in the previous section was encoded into an IBM-Q quantum processor using the {\tt Qiskit} language as a component of a three-flavor oscillations circuit as described in Ref.~\cite{arguelles2019neutrino}.  The probabilities for neutrino oscillations were calculated using a quantum simulator and found to be in agreement with theoretical expectations within available numerical precision.  Results of simulations of the original circuit from  Ref.~\cite{arguelles2019neutrino} built from two CNOT and six U3 gates with six fitted free parameters, and this circuit based on three anti-controlled unitary operations and two CNOT gates with three physical parameters, are effectively indistinguishable.  These simulation results are shown as circles and crosses on Fig.~\ref{fig:ThreeFlavor}, and on both plots they fall exactly on the theoretical oscillation probability predictions.  This validates our approach to re-creating the PMNS operation in a physically motivated way, and in an ideal universal quantum computer we would expect perfect reproduction of neutrino mixing phenomenology in the computational Hilbert space.

The circuit was then evaluated on a quantum processor, using two qubits of the {\tt IBM-Q Santiago} 5-qubit machine and 8192 evaluations (the maximum number avaialble on the {\tt Qiskit} interface).  A method was developed and presented in the Appendix of Ref.~\cite{arguelles2019neutrino} to stasticially correct for qubit readout errors by evaluating the oscillation probability at $L/E$=0 where no oscillation is expected, and using this to extract effective error rates on each qubit. This protocol was applied using the measured single-qubit error rates at $L/E=0$ of  12.5\% and 19.6\% respectively.  Repeated evaluation of the zero-distance oscillation probabilities demonstrated stable error rates within a few percent. The extracted oscillation probabilities agree qualitatively well with data for both new and old PMNS representations.  Results from the two ciruits are shown in Figs.~\ref{fig:ThreeFlavor}, left and right, respectively.

While strong qualitative agreement with the results of the simulator is observed, notably both circuits slightly under-perform the accuracy demonstrated in Ref.~\cite{arguelles2019neutrino}.  We note that the {\tt IBM-Q Santiago} computer reports somewhat higher levels of noise and yields larger bit-flip probabilities used in statistical error correction than the 5 qubit {\tt IBM-Q Yorktown} used for the original evaluation. We suspect this to be the cause of the slight observed performance reduction.  Since {\tt IBM-Q Yorktown}  has been retired we were not able to validate the hypothesis.

\begin{figure}[t]
    \centering
    \includegraphics[width=0.70\textwidth]{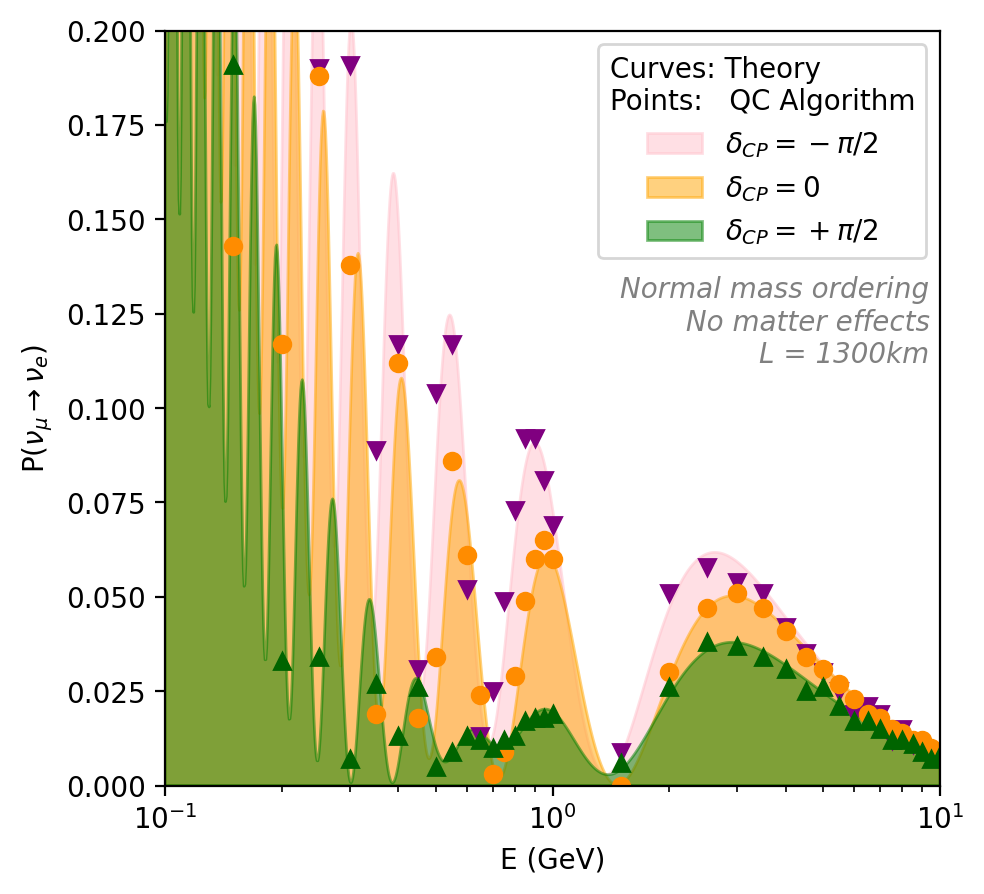} 
    \caption{Comparison of neutrino oscillation probability calculated with maximal positive and negative CP violation in absence of matter effects using the quantum algorithm developed in this work. The baseline has been chosen to be 1300~km, representative of the DUNE experiment, and neutrino masses and mixing angles match those in Ref. ~\cite{particle2020review}}
    \label{fig:CPV}
\end{figure}

The new PMNS circuit developed in this work advances beyond previous capabilities by allowing for the incorporation of neutrino CP violation.  To demonstrate proper treatment of CP-violating neutrino oscillations we have simulated the oscillation probability from muon to electron neutrinos over a baseline of 1300~km on the IBM-Q simulator, for maximal positive, zero, and maximal negative CP violation. These conditions are chosen to reflect a simplified configuration of the DUNE experiment~\cite{,abi2020deep,chatterjee2017probing}, though notably calculation here does not include the phenomenologically important effects of neutral current interactions with matter during transport. Strong agreement between simulation outputs and calculations are observed for all energies and all values of $\delta_{CP}$.

The effect of CP-violation is a small correction to the neutrino oscillation probability at these energies and baselines, and is experimentally observable primarily in the $\nu_\mu\rightarrow\nu_e$ channel which has an small amplitude (especially compared with the near maximal oscillation amplitude in the $\nu_\mu\rightarrow\nu_\tau$ channel, as tested above).  Since investigation of CP violation in vacuum neutrino oscillations requires probing $\sim$2\% changes on $\sim$5\% oscillation effects, which is beyond the precision of the IBM-Q system for 3-flavor oscillations shown in Fig~\ref{fig:ExampleManipulations}, it is unfortunately beyond the precision of the existing IBM-Q machines to accurately perform useful quantum calculations of CP violating oscillations, given what is known about the neutrino mixing matrix. Nevertheless, validity of the algorithm can be robustly established using the IBM-Q quantum simulator.  The results of this calculation compared against theoretical three-flavor CP violating neutrino oscillation probabilities is shown in Fig.~\ref{fig:CPV}.  Given a quantum computer with suitably low noise, this circuit will provide an accurate calculation of neutrino oscillations in three flavors with non-zero $\delta_{CP}$, as required.

\section{$2^N$ dimensional PMNS matrices on an $N$-qubit Hilbert space}

Assuming all permutations are available, any base sub-rotation can be used to seed the process of constructing all sub-rotations needed for the PMNS matrix.  Using the general methods of Ref.~\cite{barenco1995elementary} it is possible to build an n-qubit version of R$_{12}$ iteratively from the two-qubit version, providing a generalized N-dimensional $R^{12}(\theta_{12},\delta)$ base sub-rotation to use in construction of higher dimensional PMNS sub-matrices,
\begin{eqnarray}
R^{12}(\theta,\delta)=X^{\varotimes\label{eq:basesub} n}CU3^{[0,1,2...n-3]}_{n-1}(\theta,-\delta,\delta)CX^{[0,1,2...n-3]}_{n-2}CU3^{n-2}_{n-1}(-\theta,-\delta,\delta) \times\\
CX^{[0,1,2...n-3]}_{n-2}CU3^{n-2}_{n-1}(\theta,-\delta,\delta)X^{\varotimes n},\nonumber
\end{eqnarray}
this circuit involves a general multi-controlled unitary gate $CU3^{[0,1,2...n-2]}_{n-1}(\theta,\delta,-\delta)$ which may not be available on a all quantum architectures.  In the event that such a gate is unavailable, this gate itself can be obtained iteratively from the single-controlled $U3$ gate via the relation,
\begin{eqnarray}
CU3^{[0,1,2...n-3]}_{n-1}(\theta,-\delta,\delta)=\\CU3^{[0,1,2...n-4]}_{n-2}(\theta,-\delta,\delta)CX^{[0,1,2...n-4]}_{n-3}CU3^{n-3}_{n-2}(-\theta,-\delta,\delta) \times\nonumber\\
CX^{[0,1,2...n-4]}_{n-3}CU3^{n-3}_{n-2}(\theta,-\delta,\delta).\nonumber\end{eqnarray}
For the subset of rotations where no CP violating phase is needed, or in systems with CP conservation, the following simplified construction will be more efficient in terms of gate multiplicity,
\begin{eqnarray}
R^{12}(\theta,0)=[X^{\varotimes n-1} \varotimes\label{eq:basesubd0}\\ U3(-\theta/2,0,0)]CX^{[...]}_{n-1}[I^{\varotimes n-1} \varotimes U3(\theta,0,0)]\nonumber\\
\times CX^{[...]}_{n-1}[X^{\varotimes n-1} \varotimes U3(-\theta/2,0,0)].\nonumber
\end{eqnarray}
The equivalence of this circuit with Eq.~\ref{eq:basesub} is proven in Appendix A.

To extend the protocol demonstrated earlier to $n$ qubits, we require the permutation operators $(ij,\mathbf{n})$ for all $i,j$ in terms of quantum circuits on the $n$ qubit system.  For this purpose, it suffices to obtain circuits for each permutation ($i,i+1;\mathbf{n}$) for each $i\leq 2^n-1$. The remaining circuits can be obtained by repeated application of the expression,
\begin{equation}
    (iq;\mathbf{n})(rj;\mathbf{n})(qr;\mathbf{n})(rj;\mathbf{n})(iq;\mathbf{n})=(ij;\mathbf{n}).\label{eq:permformula}
\end{equation}
We introduce controlled and anti-controlled permutation operations, which are n-qubit controlled unitary operations, with control qubit $m$ acting on the remaining $N-1$ qubits, as,
\begin{eqnarray}
C(ij;\mathbf{n-1})^m_{[...]}\quad \quad \slashed{C}(ij;\mathbf{n-1})^m_{[...]}.
\end{eqnarray}
Our protocol will be to construct these $n$-dimensional controlled operations in terms similar controlled operations of lower dimensionality, leading to an iterative recipe for obtaining the group of permutation circuits $(ij,\mathbf{n})$ on an n-qubit space.

We will find the permutations of dimension $2^n$ in three classes, distinguished by their action on the first qubit,
\begin{eqnarray}
    I:&(ij;\boldsymbol{n})|i;\boldsymbol{n}\rangle=|0\rangle\varotimes|j;\boldsymbol{n-1}\rangle&,\\& \nonumber (ij;\boldsymbol{n})|j;\boldsymbol{n}\rangle=|0\rangle\varotimes|i;\boldsymbol{n-1}\rangle;\\ \nonumber
    \\ 
    II:&(i+2^{n-1},j+2^{n-1};\boldsymbol{n})|i+2^{n-1};\boldsymbol{n}\rangle=|1\rangle\varotimes |j;\boldsymbol{n-1}\rangle, \\ & \nonumber (i+2^{n-1},j+2^{n-1};\boldsymbol{n})|j+2^{n-1};\boldsymbol{n}\rangle=|1\rangle\varotimes |i;\boldsymbol{n-1}\rangle\\ \nonumber
    \\
    III: &\mathrm{Neither\,I\,or\,II}.&
\end{eqnarray}

By examination, we see that the permutation operators $(i,i+i;\mathrm{n})$ for $i<2^{n-1}$ are in Class I, and for $i>2^{n-1}$ are in Class II. This leaves only one such permutation operator in Class III, which is $(2^{n-1},2^{n-1}+1;\mathrm{n})$.

Consideration of the action of the relevant controlled and anti-controlled unitary operations allows us to identify the elements in Class I and Class II as anti-controlled and controlled lower-dimensional permutation operators, respectively, as,
\begin{eqnarray}
I: &(i,i+1;\boldsymbol{n})=\slashed{C}(i,i+1;\boldsymbol{n-1})^0_{[...]}&i<2^{n-1}\\
II:&(i,i+1;\boldsymbol{n})= C(i-2^{n-1},i-2^{n-1}+1;\boldsymbol{n-1})^0_{[...]}&i>2^{n-1}
\end{eqnarray}
The permutation still to be found exchanges basis states $|0111...\rangle$ with $|1000...\rangle$. We can obtain this final permutation by noting that,
\begin{equation}
C(2^{n-2},2^{n-2}+1;\boldsymbol{n-1})^{n-1}_{[...]}=(2^{n-1} \ 2^{n-1}+2;\boldsymbol{n}).
\end{equation}
and then applying Eq.~\ref{eq:permformula} to obtain,
\begin{eqnarray}
III:&(2^{n-1},2^{n-1}+1;\boldsymbol{n})=\left(2^{n-1}+1, 2^{n-1}+2; \boldsymbol{n}\right) \times \nonumber \\  & C\left(2^{n-2},2^{n-2}+1;\boldsymbol{n-1}\right)^{n-1}_{[...]}\left(2^{n-1}+1,2^{n-1}+2; \boldsymbol{n}\right).
\end{eqnarray}
Thus all permutation operators emerge as controlled or anti-controlled versions of lower dimensional operations.  

The iterative recipe is completed by finding the permutation operations on one qubit. Only one permutation operation exists on the one-qubit system, which is $(0,1;\mathbf{1})$.  This operation switches $|0\rangle$ for $|1\rangle$, so we identify it as the simple single-qubit X gate, 
\begin{equation}
    (1,2;\mathbf{1})=\begin{bmatrix} 0 &1 \\ 1 & 0 \end{bmatrix}=X.
\end{equation}
All higher dimensional permutation circuits can be generated from multi-controlled X gates. 

As an example, to generate the two-qubit PMNS matrix we must find three distinct permutation operations, $(1,2;\mathbf{2})$ in Class I, $(2,3;\mathbf{2})$ in Class III, and $(3,4;\mathbf{2})$ in Class II.  These are given according to the above recipe by,
\begin{eqnarray}
(1,2;\mathbf{2})&=&\slashed{C} X^0_{1},\\
(2,3;\mathbf{2})&=&C X^0_{1} C X^{1}_{0}C X^0_{1},\\
(3,4;\mathbf{2})&=&C X^0_{1}.
\end{eqnarray}
The remaining permutations needed to generate all PMNS matrix elements are given by Eq.~\ref{eq:permformula}. It can be verified that these sub-rotation circuits so constructed can be simplified into the two-qubit circuits demonstrated in Sec.~\ref{sec:PMNS3D}, as shown in Appendix B.

\begin{figure}[t]
    \centering
    \includegraphics[width=0.99\textwidth]{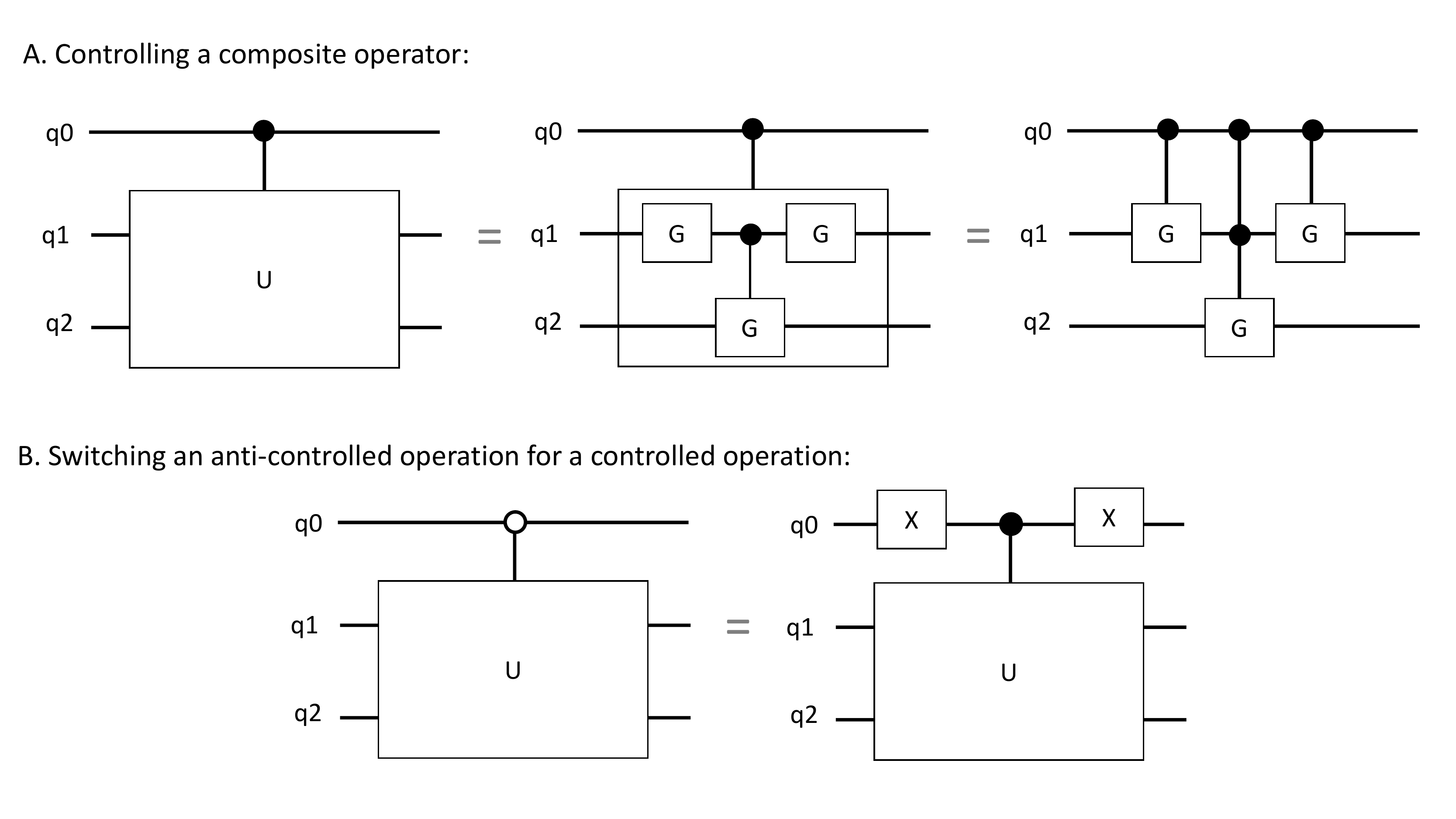} 
    \caption{Basic manipulations required to generate n-dimensional mixing circuit in terms of only mutli-controlled X and single-qubit U3 gates. A: composite controlled operations; B: negation of an anti-control qubit.}
    \label{fig:ExampleManipulations}
\end{figure}

For higher dimensional systems, this recipe will require controlled versions of multi-qubit operations; these are generated in the standard way, by controlling each gate in series, per Fig.~\ref{fig:ExampleManipulations}, top.  All anti-control operations can also be substituted for control operations by application of pairs of X gates to the control qubit, as shown in Fig.~\ref{fig:ExampleManipulations}, bottom. Thus the result of this procedure for generating all possible permutation operations is a set of quantum n-qubit circuits involving only multi-controlled X gates.   

With all permutations obtained, the PMNS matrix operation is assembled using Eq.~\ref{eq:BaseSubrotPermut} with the n-qubit base sub-rotatation $R_{12}$ of Eq.~\ref{eq:basesub} and the appropriate permutation operations to generate each $R_{ij}$, which are combined in series to yield the full PMNS matrix.   Finally, with the PMNS sub-circuit generated for (up to) $2^n$ physical neutrinos, the neutrino oscillation simulation circuit can then be constructed as in Ref.~\cite{arguelles2019neutrino}, using single qubit U1 gates to apply time evolution in the mass basis between two applications of the PMNS operation.  This completes the construction of a quantum algorithm for direct quantum simulation of oscillations between arbitrary numbers of mixed, massive neutrinos.

\section{Conclusions}

We have obtained a complete protocol for quantum simulation of oscillations between $2^n$ arbitrarily mixed neutrinos with arbitrary masses, including CP-violation, on an n-qubit quantum computer.  

Construction of the $2^n$ dimensional PMNS operation on $n$ qubits proceeds iteratively, yielding a circuit constructed from elementary single qubit U3 operations and multi-controlled X gates.  Each  sub-rotation of the PMNS matrix  $R_{ij}(\theta_{ij},\delta_{ij})$ in the standard parameterization is obtained by applying permutation operations to a base sub-rotation $R_{12}(\theta_{ij},\delta_{ij})$.  These sub-rotations applied in series yield the PMNS sub-circuit. Neutrino time evolution can then be implemented as in Ref.~\cite{arguelles2019neutrino} in the mass basis, using single-qubit U1 gates.   Quantum simulations of the three-neutrino oscillation system both with and without CP violation is used as a validation of the approach, reproducing the expected theoretical oscillation probabilities on publicly available IBM-Q quantum computers.  

The protocol we have presented improves substantially upon the empirical approach used Ref.~\cite{arguelles2019neutrino} in several important ways. The number of free parameters in the circuit equals the number of physical mixing parameters; CP violation is incorporated; and the protocol can be applied to systems of arbitrarily high dimensionality.  Finally, the form of the circuit used is determined by rigorous mathematical and physical reasoning rather than trial-and-error, as in the previous work. This improved mathematical rigor coupled with substantially expanded physical realm of applicability represents an important step forward in mapping the calculation of neutrino oscillation probabilities onto a problem well posed for solution on a quantum processor.  

Finally, we note that the general approach to the problem of simulation of neutrino oscillations on $2^{n}$ qubits may be applied to a wide variety of other quantum problems.  The process of decomposing the system into a basis of energy Eigenstates via a unitary operation, which can be without loss of generality expressed in terms of a series of sub-rotations, time evolving them in the energy basis, and then rotating back to the physically observable basis, is a completely generic way to implement time evolution of any quantum system.  While neutrino oscillations have been our test case for this work, arbitrary quantum systems may be similarly treated, using the protocol presented.



\section*{Appendix A: Proof of the $\delta_{CP}=0$ simplified base sub-rotation from the general case\label{sec:AppendixA}}

In this appendix we prove that Eq.~\ref{eq:basesubd0} is equivalent to Eq.~\ref{eq:basesub} with $\delta=0$. First we note that Eq.~\ref{eq:basesubd0} can be written as,
\begin{eqnarray}
R^{12}(\theta ,0)=X^{\varotimes n}[I^{\varotimes n-1} \varotimes U3(\theta /2,0,0)X]CX^{[0,1,2...n-2]}_{n-1}[I^{\varotimes n-1} \varotimes U3(\theta,0,0)X] \times \\ 
CX^{[0,1,2...n-2]}_{n-1}[I^{\varotimes n-1} \varotimes U3(\theta /2,0,0)]X^{\varotimes n},\nonumber
\end{eqnarray}
or equivalently,
\begin{eqnarray}
= X^{\varotimes n}[I^{\varotimes n-1} \varotimes U3(\theta /2,0,0)][I^{n-1} \varotimes X]CX^{[0,1,2...n-2]}_{n-1}[I^{\varotimes n-1} \varotimes U3(\theta,0,0)] \\
 \times [I^{\varotimes n-1} \varotimes X]CX^{[0,1,2...n-2]}_{n-1}[I^{\varotimes n-1} \varotimes U3(\theta /2,0,0)]X^{\varotimes n}. \nonumber    
\end{eqnarray}
We can decompose this into a product of sub-rotations and permutations, as follows,
\begin{eqnarray}
=(1,2^n)(2,2^n-1)...(2^{n-1},2^{n-1}+1)R^{12}(-\theta/4,0)R^{34}(-\theta/4,0)...R^{2^n-1,2^n}(-\theta/4,0) \times\\
(12)(34)...(2^n-1,2^n)(2^n-1,2^n)R^{12}(-\theta/2,0)R^{34}(-\theta/2,0)...R^{2^n-1,2^n}(-\theta/2,0) \times\nonumber\\
(12)(34)...(2^n-1,2^n)(2^n-1,2^n)R^{12}(-\theta/4,0)R^{34}(-\theta/4,0)...R^{2^n-1,2^n}(-\theta/4,0)\times\nonumber\\
(1,2^n)(2,2^n-1)...(2^{n-1},2^{n-1}+1).\nonumber
\end{eqnarray}
Making the obvious cancellations and noting that $(ij)R^{ij}(\theta,0)(ij)=R^{ji}(\theta,0)=R^{ij}(-\theta,0)$ on the middle sub-rotations, we have,
\begin{eqnarray}
=(1,2^n)(2,2^n-1)...(2^{n-1},2^{n-1}+1)R^{12}(-\theta/4,0)R^{34}(-\theta/4,0)...R^{2^n-1,2^n}(-\theta/4,0) \times\\
R^{12}(\theta/2,0)R^{34}(\theta/2,0)...R^{2^n-1,2^n}(-\theta/2,0) R^{12}(-\theta/4,0)R^{34}(-\theta/4,0)...\times\nonumber\\
R^{2^n-1,2^n}(-\theta/4,0)(1,2^n)(2,2^n-1)...(2^{n-1},2^{n-1}+1).\nonumber
\end{eqnarray}
Combining sub-rotations,
\begin{eqnarray}
(1,2^n)(2,2^n-1)...(2^{n-1},2^{n-1}+1)R^{2^n-1,2^n}(-\theta,0)(1,2^n)(2,2^n-1)...(2^{n-1},2^{n-1}+1)=\\
(1,2^n)(2,2^n-1)R^{2^n-1,2^n}(-\theta,0)(1,2^n)(2,2^n-1)=\nonumber\\
R^{21}(-\theta,0)=R^{12}(\theta,0).\nonumber
\end{eqnarray}
Which is equivalent to Eq.~\ref{eq:basesub}, as was to be proven.

\section*{Appendix B: Extraction of two-qubit circuits from the $n$-qubit recipe}
Using the protocol developed and the equations for our generators, we can reproduce the sub-rotations presented in section~\ref{sec:PMNS3D}. As $R^{12}(\theta,\delta)$ is our base sub-rotation, we need only obtain the the other 5. For $R^{23}$ we have,
\begin{equation}
R^{23}(\theta,\delta)=(12)(23)R^{12}(\theta,\delta)(23)(12).
\end{equation}
Noting that the $(23;\boldsymbol{2})$ we obtained above is equivalent to the SWAP gate, this gives us,
\begin{equation}
\scalebox{1.0}{
\Qcircuit @C=1.0em @R=0.2em @!R {
	 	\nghost{ {q}_{0} :  } & \lstick{ {q}_{0} :  } & \ctrlo{1} & \qswap & \ctrlo{1} & \qswap & \ctrlo{1} & \qw & \qw\\ 
	 	\nghost{ {q}_{1} :  } & \lstick{ {q}_{1} :  } & \targ & \qswap \qwx[-1] & \gate{\mathrm{U_3}\,(\mathrm{-2\theta,\delta,-\delta})} & \qswap \qwx[-1] & \targ & \qw & \qw\\ 
}}
\end{equation}
For $R^{13}$ we have,
\begin{equation}
R^{13}(\theta,\delta)=(23)R^{12}(\theta,\delta)(23).
\end{equation}
Hence $R^{13}$ is, 
\begin{equation}
\scalebox{1.0}{
\Qcircuit @C=1.0em @R=0.2em @!R {
	 	\nghost{ {q}_{0} :  } & \lstick{ {q}_{0} :  } & \qswap & \ctrlo{1} & \qswap & \qw & \qw\\ 
	 	\nghost{ {q}_{1} :  } & \lstick{ {q}_{1} :  } & \qswap \qwx[-1] & \gate{\mathrm{U_3}\,(\mathrm{-2\theta,\delta,-\delta})} & \qswap \qwx[-1] & \qw & \qw\\ 
 }}
\end{equation}
Next $R^{14}$,
\begin{equation}
R^{14}(\theta,\delta)=(34)(23)R^{12}(\theta,\delta)(23)(34).
\end{equation}
Writing (23) in terms of CNOT gates and taking advantage of the cancellation with (34) this affords us, we obtain,
\begin{equation}
\scalebox{1.0}{
\Qcircuit @C=1.0em @R=0.2em @!R {
	 	\nghost{ {q}_{0} :  } & \lstick{ {q}_{0} :  } & \targ & \ctrl{1} & \ctrlo{1} & \ctrl{1} & \targ & \qw & \qw\\ 
	 	\nghost{ {q}_{1} :  } & \lstick{ {q}_{1} :  } & \ctrl{-1} & \targ & \gate{\mathrm{U_3}\,(\mathrm{-2\theta,\delta,-\delta})} & \targ & \ctrl{-1} & \qw & \qw 
}}
\end{equation}
Noting that the CNOT gates on either side of $R^{12}$ commute with it, $R^{14}$ simplifies to,
\begin{equation}
\scalebox{1.0}{
\Qcircuit @C=1.0em @R=0.2em @!R { \\
	 	\nghost{ {q}_{0} :  } & \lstick{ {q}_{0} :  } & \targ & \ctrlo{1} & \targ & \qw & \qw\\ 
	 	\nghost{ {q}_{1} :  } & \lstick{ {q}_{1} :  } & \ctrl{-1} & \gate{\mathrm{U_3}\,(\mathrm{-2\theta,\delta,-\delta})} & \ctrl{-1} & \qw & \qw\\ 
}}
\end{equation}
$R^{24}$ is:
\begin{equation}
R^{24}(\theta,\delta)=(12)(34)(23)R^{12}(\theta,\delta)(23)(34)(12),
\end{equation}
which is equal to,
\begin{equation}
\scalebox{1.0}{
\Qcircuit @C=1.0em @R=0.2em @!R { \\
	 	\nghost{ {q}_{0} :  } & \lstick{ {q}_{0} :  } & \ctrlo{1} & \ctrl{1} & \qswap & \ctrlo{1} & \qswap & \ctrl{1} & \ctrlo{1} & \qw & \qw\\ 
	 	\nghost{ {q}_{1} :  } & \lstick{ {q}_{1} :  } & \targ & \targ & \qswap \qwx[-1] & \gate{\mathrm{U_3}\,(\mathrm{-2\theta,\delta,-\delta})} & \qswap \qwx[-1] & \targ & \targ & \qw & \qw\\ 
}}.
\end{equation}
The combination of the control and anti-control bits on this circuit reduces $R^{24}$ to,
\begin{equation}
\scalebox{1.0}{
\Qcircuit @C=1.0em @R=0.2em @!R { \\
	 	\nghost{ {q}_{0} :  } & \lstick{ {q}_{0} :  } & \qw & \qswap & \ctrlo{1} & \qswap & \qw & \qw & \qw\\ 
	 	\nghost{ {q}_{1} :  } & \lstick{ {q}_{1} :  } & \gate{\mathrm{X}} & \qswap \qwx[-1] & \gate{\mathrm{U_3}\,(\mathrm{-2\theta,\delta,-\delta})} & \qswap \qwx[-1] & \gate{\mathrm{X}} & \qw & \qw\\ 
}}
\end{equation}
Finally for $R^{34}$, 
\begin{equation}
R^{34}(\theta,\delta)=(23)(12)(34)(23)R^{12}(\theta,\delta)(23)(34)(12)(23).
\end{equation}
Writing the SWAP gate again as a combination of CNOT gates and using this to cancel out the (34)'s, we obtain,
\begin{equation}
\scalebox{1.0}{
\Qcircuit @C=1.0em @R=0.2em @!R {
	 	\nghost{ {q}_{0} :  } & \lstick{ {q}_{0} :  } & \ctrl{1} & \targ & \ctrl{1} & \ctrlo{1} & \targ & \ctrl{1} & \ctrlo{1} & \ctrl{1} & \targ & \ctrlo{1} & \ctrl{1} & \targ & \ctrl{1} & \qw & \qw\\ 
	 	\nghost{ {q}_{1} :  } & \lstick{ {q}_{1} :  } & \targ & \ctrl{-1} & \targ & \targ & \ctrl{-1} & \targ & \gate{\mathrm{U_3}\,(\mathrm{-2\theta,\delta,-\delta})} & \targ & \ctrl{-1} & \targ & \targ & \ctrl{-1} & \targ & \qw & \qw\\ 
}}
\end{equation}
Taking advantage of combinations of control and anti-control bits, noting that we can move X gates across control or anti-control bits by flipping them to their opposites, and observing that X gates can be moved across the target bits of CNOT gates with impunity, this $R^{34}$ to,
\begin{equation}
\scalebox{1.0}{
\Qcircuit @C=1.0em @R=0.2em @!R { \\
	 	\nghost{ {q}_{0} :  } & \lstick{ {q}_{0} :  } & \gate{\mathrm{X}} & \ctrlo{1} & \gate{\mathrm{X}} & \qw & \qw\\ 
	 	\nghost{ {q}_{1} :  } & \lstick{ {q}_{1} :  } & \qw & \gate{\mathrm{U_3}\,(\mathrm{-2\theta,\delta,-\delta})} & \qw & \qw & \qw\\ 
}}
\end{equation}
completing the set of two-qubit sub-rotations.

\section*{Acknowledgements}
We acknowledge the use of IBM Quantum services for this work. The views expressed are those of the authors, and do not reflect the official policy or position of IBM or the IBM Quantum team.  We thank Carlos Arg\"uelles, Raquel Castillo Fernandez, Jonathan Asaadi, Krishan Mistry and Grant Parker for their comments on the manuscript and valuable insights.  BJPJ is supported by  the Department of Energy under Early Career Award number DE-SC0019054 and the National Science Foundation under award number 1913607.

\bibliography{main}

\end{document}